# Light-Shift Imbalance Induced Blockade of Collective Excitations Beyond the Lowest Order

*M.S. Shahriar, P. Pradhan, G.S. Pati, V.Gopal, and K. Salit*
*EECS Department, Northwestern University,*
*Evanston, IL 60208*

## Abstract

Current proposals focusing on neutral atoms for quantum computing are mostly based on using single atoms as quantum bits (qubits), while using cavity induced coupling or dipole-dipole interaction for two-qubit operations. An alternative approach is to use atomic ensembles as quantum bits. However, when an atomic ensemble is excited, by a laser beam matched to a two-level transition (or a Raman transition) for example, it leads to a cascade of many states as more and more photons are absorbed[1]. In order to make use of an ensemble as a qubit, it is necessary to disrupt this cascade, and restrict the excitation to the absorption (and emission) of a single photon only. Here, we show how this can be achieved by using a new type of blockade mechanism, based on the light-shift imbalance (LSI) in a Raman transition. We describe first a simple example illustrating the concept of light shift imbalance induced blockade (LSIIB) using a multi-level structure in a single atom, and show verifications of the analytic prediction using numerical simulations. We then extend this model to show how a blockade can be realized by using LSI in the excitation of an ensemble. Specifically, we show how the LSIIB process enables one to treat the ensemble as a two level atom that undergoes fully deterministic Rabi oscillations between two collective quantum states, while suppressing excitations of higher order collective states.

## A. Introduction

Many different technologies are currently being pursued for realizing a quantum computer (QC). One of the most promising approaches involve the use of neutral atoms. This approach is particularly attractive because, in principle, it is possible to achieve very long decoherence times and very high fidelities when using neutral atoms. Current proposals for quantum computing focusing on neutral atoms are based on using single atoms as quantum bits, often while using cavity induced coupling or dipole-dipole interaction for two-qubit operations. However, given the degree of difficulties encountered in isolating and controlling single atoms, this process has proven very difficult to realize, especially on a large scale. An alternative approach is to use atomic ensembles as quantum bits. However, when an atomic ensemble is excited, by a laser beam matched to a two-level transition (or a Raman transition) for example, it leads to a cascade of many states as more and more photons are absorbed[1, 2, 3]. In order to make use of an ensemble as a qubit, it is necessary to disrupt this cascade, and restrict the excitation to the absorption (and emission) of a single photon only. In principle, this can be achieved through the use of the so-called dipole blockade, which can be particularly efficient if Rydberg transitions are used[4, 5].

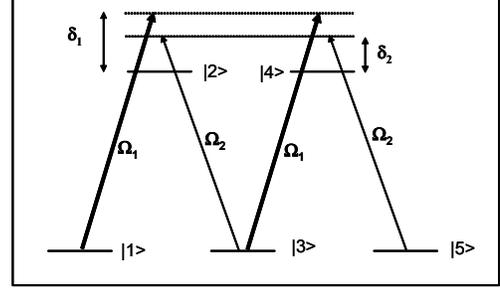

**Figure 1:** *Schematic illustration of a five level system, illustrating the process of LSIIB. See text for details.*

Dipole blockades generally occur between individual atoms within an ensemble. In order to make use of this blockade mechanism in a manner that is consistent with a quantum computing architecture, it is necessary to control the distribution of inter-atomic distances between each pair of atoms in the ensemble in a precise manner Furthermore, in order to achieve long decoherence times, it is necessary to make use of dipole-blockades based on spin-spin coupling, which is necessarily much weaker than the optical dipole-dipole coupling.

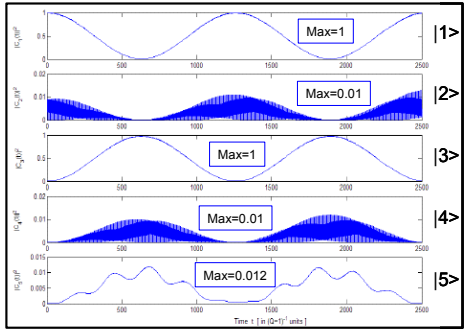

**Figure 3:** *Populations of the five level system, starting with all the atoms in level 1. See text for details.*

Here, we show that a new type of blockade mechanism, based on the light-shift imbalance in a Raman transition, can overcome these limitations. The resulting system does not impose any constraint on the distribution of inter-atomic distance within an ensemble. Furthermore, no dipole-dipole coupling is necessary, so that a relatively low density system can be used.

In what follows, we describe first a simple example illustrating the concept of light shift imbalance induced blockade (LSIIB) using a multi-level structure in a single atom, and show verifications of the analytic prediction using numerical simulations. We then extend this model to show how a blockade can be realized by using LSI in the excitation of an ensemble. Specifically, we show how the LSIIB process enables one to treat the ensemble as a two level atom that undergoes fully deterministic Rabi oscillations between two collective quantum states, while suppressing excitations of higher order collective states. We then show how this transition can be used to realize a quantum bit (qubit) embodied by the ensemble. Using multiple energy levels inside each atom, the LSIIB enables the transfer of quantum information between neighboring ensembles, as well as the realization of a CNOT gate. In effect, this represents a generalization of the so-called Pellizari scheme for quantum information processing[6]. Furthermore, the LSIIB can be used to link two separate quantum computers (QC), by transferring the quantum state of any ensemble qubit in one QC to any ensemble qubit in another QC. In a separate paper, we discuss details of these quantum computation and communication protocols, offer practical ways to implement this scheme, and propose experiments to demonstrate the feasibility of these schemes[7].

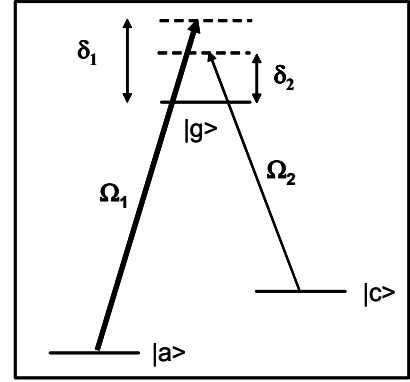

**Figure 2:** *Schematic illustration of a three level transition in each atom in an ensemble.*

The significance of the LSIIB process can be summarized as follows: (a) It can be used to realize a deterministic quantum bit encoded in the collective-excitation states of an atomic ensemble. (b) Along with a moderate-Q cavity, it can be used to realize a two-qubit gate (e.g., a C-NOT gate) between two ensemble-based qubits. (c) It can be used to transport, deterministically, the quantum state of an ensemble qubit from one location to another separated by macroscopic distances, and (d) It can be used to establish a quantum-link between two ensembles-and-cavity based quantum computers. The scheme proposed here and expanded further in reference 7 therefore offers a robust technique for realizing a quantum internet without using the single-atom and super-cavity based approaches[8,9,10].

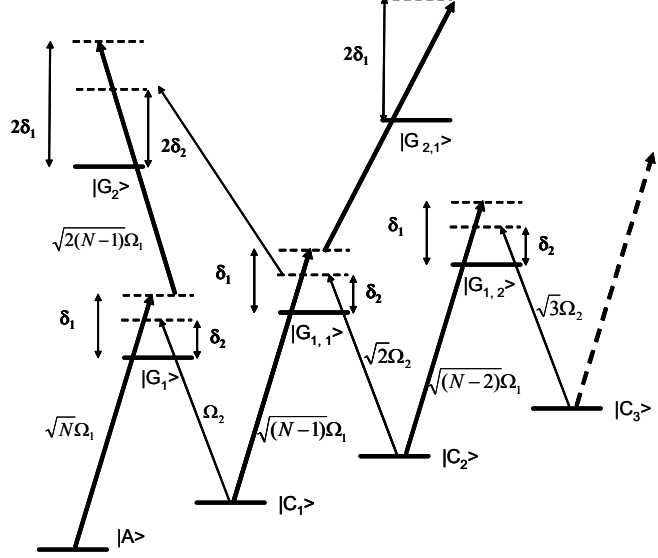

**Figure 4:** *Schematic illustration of the relevant collective states and the corresponding transition rates. See text for details*

### B. Light Shift Imbalance Induced Blockade In a Single Atom: An Illustrative Example

In order to illustrate the basic mechanism that underlies the LSIIB, it is convenient to consider first a simple example of a set of five levels inside a single atom, as shown in Figure 1. Under the rotating wave approximation (RWA), and the rotating wave transformation (RWT), the Hamiltonian (with $\hbar = 1$) describing this interaction is given, in the bases of $\{|1>,|2>,|3>,|4>,|5>\}$, by:

$$H = \begin{bmatrix} \Delta/2 & \Omega_1/2 & 0 & 0 & 0 \\ \Omega_1/2 & -\delta & \Omega_2/2 & 0 & 0 \\ 0 & \Omega_2/2 & -\Delta/2 & \Omega_1/2 & 0 \\ 0 & 0 & \Omega_1/2 & -(\delta+\Delta) & \Omega_2/2 \\ 0 & 0 & 0 & \Omega_2/2 & -3\Delta/2 \end{bmatrix} \quad (1)$$

where we have defined $\delta \equiv (\delta_1 + \delta_2)/2$ and $\Delta \equiv (\delta_1 - \delta_2)$. Under the conditions that $\delta >> \Delta$, $\delta >> \Omega_1$, and $\delta >> \Omega_2$, one can eliminate the optically excited states |2> and |4> adiabatically. The effective Hamiltonian in the bases of $\{|1>,|3>,|5>\}$ is then given by[11,12,13,14]:

$$\tilde{H} = \begin{bmatrix} (\Delta/2 + \varepsilon_1) & \Omega_R/2 & 0 \\ \Omega_R/2 & [-\Delta/2 + (\varepsilon_1 + \varepsilon_2)] & \Omega_R/2 \\ 0 & \Omega_R/2 & (-3\Delta/2 + \varepsilon_2) \end{bmatrix} \quad (2)$$

where $\Omega_R \equiv (\Omega_1 \Omega_2)/2\delta$ is the Raman-Rabi frequency, and $\varepsilon_j \equiv \Omega_j^2/4\delta$ is the light shift due to $\Omega_j$ (*j=1,2*). Note that the levels |1>, |3> and |5> are light shifted by different amounts. In general, this Hamiltonian describes a process wherein populations can oscillate between the states |1>, |3> and |5>, with the maximum amplitude in each level being determined by the relative values of the parameter.

Consider now the case where $\Omega_2 \ll \Omega_1$. Furthermore, assume that $\Delta = \varepsilon_2$. Under this condition, the Raman coupling between |1> and |3> become resonant, while the Raman coupling between |3> and |5> becomes detuned by $-(\varepsilon_1 + \varepsilon_2)$. Explicitly, this can be seen by subtracting an energy $(\varepsilon_1 + \Delta/2)$, to give:

$$\widetilde{\widetilde{H}} = \begin{bmatrix} 0 & \Omega_R/2 & 0 \\ \Omega_R/2 & 0 & \Omega_R/2 \\ 0 & \Omega_R/2 & -(\varepsilon_1 + \varepsilon_2) \end{bmatrix} \quad (3)$$

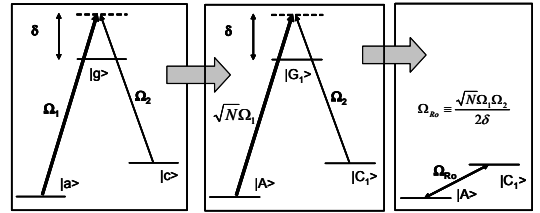

**Figure 5:** *Summary of the LSIIB process in an ensemble. See text for details.*

In the limit of $\varepsilon_2 \to 0$, $\Delta = 0$, the detuning for the |3> to |5> Raman coupling is simply $-\varepsilon_1$. For $\Omega_R \ll \varepsilon_1$ (which is the same condition as $\Omega_2 \ll \Omega_1$), the coupling to level |5> can be ignored. The net result is that the system will oscillate between |1> and |3>. Note that this result is due to the fact that levels |1> and |3> get light shifted by nearly the same amounts, thus remaining resonant for the Raman transition, while level |5> sees virtually no light shift. Thus, the excitation to level |5> is essentially blockaded by the imbalance in the light shifts.

We have verified this result using numerical simulations, as shown in figure 2. These are calculated for the following parameter values: $\Delta = 0$, $\Omega_2/\Omega_1 = 0.1$, and $\delta/\Omega_1 = 10$. Analytically, the residual populations in levels |2> and |4> are expected to be of the order of $(\Omega_1/\delta)^2$ and that of level |5> is expected to be of the order of $(\Omega_R/\varepsilon_1)^2$, which in turn is of the order of $(\Omega_2/\Omega_1)^2$ and are consistent with the values seen here. These excitations can be suppressed further by making these ratios smaller.

**C.      LSIIB in Ensemble Excitation**

Consider next the excitation of an ensemble of N atoms. To start with, we consider each atom in the ensemble to be a three level system, as illustrated in figure 3. Using Dicke's model of collective excitation[1], we can show that the ensemble excitation can be represented as shown in figure 4. The collective states in this diagram are defined as follows:

$$|A\rangle = |a_1, a_2, \cdots, a_N\rangle; \; |G_1\rangle = \frac{1}{\sqrt{N}} \sum_{j=1}^{N} |a_1, a_2, \cdots, g_j, \cdots a_N\rangle; \; |C_1\rangle = \frac{1}{\sqrt{N}} \sum_{j=1}^{N} |a_1, a_2, \cdots, c_j, \cdots a_N\rangle$$

$$|G_2\rangle = \frac{1}{\sqrt{{}^NC_2}} \sum_{j,k(j\neq k)}^{{}^NC_2} |a_1, a_2, \cdots, g_j, \cdots, g_k, \cdots a_N\rangle]; \; |C_2\rangle = \frac{1}{\sqrt{{}^NC_2}} \sum_{j,k(j\neq k)}^{{}^NC_2} |a_1, a_2, \cdots, c_j, \cdots, c_k, \cdots a_N\rangle$$

$$|G_{1,1}\rangle = \frac{1}{\sqrt{2{}^NC_2}} \sum_{j,k(j\neq k)}^{2{}^NC_2} |a_1, a_2, \cdots, g_j, \cdots, c_k, \cdots a_N\rangle$$

$$|G_{2,1}\rangle = \frac{1}{\sqrt{Z}} \sum_{(j\neq k\neq l)}^{Z} |a_1, a_2, \cdots, g_j, \cdots, g_k, \cdots, c_l, \cdots a_N\rangle$$

$$|G_{1,2}\rangle = \frac{1}{\sqrt{Z}} \sum_{(j\neq k\neq l)}^{Z} |a_1, a_2, \cdots, g_j, \cdots, c_k, \cdots, c_l, \cdots a_N\rangle \quad [Z = 3{}^NC_3]$$

(4)

The relevant coupling rates between these collective states are also illustrated in figure 4. Note that for large detunings, the excitations to the intermediate states $|G_1\rangle$ and $|G_{1,1}\rangle$ are small, so that higher order states such as $|G_2\rangle$ and $|G_{2,1}\rangle$ can be ignored. The remaining system looks very similar to the five-level system considered in section B above. (Parenthetically at this point, note that the coupling between $|G_1\rangle$ and $|C_1\rangle$ does not scale with √N, unlike the coupling between $|A\rangle$ and $|G_1\rangle$, which scales as √N.)

We can now proceed in a manner similar to that in section B. First, in the rotating wave transformation frame, the truncated, six level Hamiltonian, in the bases of $|A\rangle$, $|G_1\rangle$, $|C_1\rangle$, $|G_{1,1}\rangle$, $|C_2\rangle$ and $|G_{1,2}\rangle$ is given by (the justification for not including the state $|C_3\rangle$ will be made by showing that the excitation to $|C_2\rangle$ can be suppressed, thus in turn making the amplitude of $|C_3\rangle$ insignificant):

$$H = \begin{bmatrix} \Delta/2 & \sqrt{N}\Omega_1/2 & 0 & 0 & 0 & 0 \\ \sqrt{N}\Omega_1/2 & -\delta & \Omega_2/2 & 0 & 0 & 0 \\ 0 & \Omega_2/2 & -\Delta/2 & (\sqrt{N-1})\Omega_1/2 & 0 & 0 \\ 0 & 0 & (\sqrt{N-1})\Omega_1/2 & -(\delta+\Delta) & \sqrt{2}\Omega_2/2 & 0 \\ 0 & 0 & 0 & \sqrt{2}\Omega_2/2 & -3\Delta/2 & (\sqrt{N-2})\Omega_1/2 \\ 0 & 0 & 0 & 0 & (\sqrt{N-2})\Omega_1/2 & -(\delta+2\Delta) \end{bmatrix} \quad (5)$$

where the detunings are defined just as before: $\delta \equiv (\delta_1 + \delta_2)/2$ and $\Delta \equiv (\delta_1 - \delta_2)$.

If the detunings are large compared to the transition rates, we can eliminate states $|G_1\rangle$, $|G_{1,1}\rangle$ and $|G_{1,2}\rangle$ adiabatically. Under this condition, the effective Hamiltonian for the three remaining states ($|A\rangle$, $|C_1\rangle$, and $|C_2\rangle$) are given by (assuming $\delta \gg \Delta$):

$$\widetilde{H} = \begin{bmatrix} \varepsilon_A + \Delta/2 & \Omega_{Ro}/2 & 0 \\ \Omega_{Ro}/2 & \varepsilon_{C1} - \Delta/2 & \sqrt{[2(N-1)/N]}\Omega_{Ro}/2 \\ 0 & \sqrt{[2(N-1)/N]}\Omega_{Ro}/2 & \varepsilon_{C2} - 3\Delta/2 \end{bmatrix} \quad (6)$$

where $\varepsilon_A$, $\varepsilon_{C1}$, and $\varepsilon_{C2}$ are the light shifts of the states $|A\rangle$, $|C_1\rangle$, and $|C_2\rangle$, respectively. and $\Omega_{Ro} \equiv (\sqrt{N}\Omega_1\Omega_2)/(2\delta)$. To first order, these light shifts are balanced, in the sense that $(\varepsilon_{C1} - \varepsilon_A) = (\varepsilon_{C2} - \varepsilon_{C1})$. This means that if the explicit two-photon detuning, $\Delta$, is chosen to make the Raman transition between $|A\rangle$ and $|C_1\rangle$ resonant (i.e., $\Delta = (\varepsilon_{C1} - \varepsilon_A)$), then the Raman transition between $|C_1\rangle$ and $|C_2\rangle$ also becomes resonant. This balance is broken when the light shifts are calculated to second order, and the blockade shift is then given by

$$\Delta_B \equiv (\varepsilon_{C2} - \varepsilon_{C1}) - (\varepsilon_{C1} - \varepsilon_A) = -(\Omega_2^4 + \Omega_1^4)/(8\delta^3).$$

With the proper choice of two-photon detuning ($\Delta = (\varepsilon_{C1} - \varepsilon_A)$) to make the Raman transition between $|A\rangle$ and $|C_1\rangle$ resonant, the effective Hamiltonian (after shifting the zero of energy, and assuming $N \gg 1$) is now given by:

$$\widetilde{H} = \begin{bmatrix} 0 & \Omega_{Ro}/2 & 0 \\ \Omega_{Ro}/2 & 0 & \Omega_{Ro}/\sqrt{2} \\ 0 & \Omega_{Ro}/\sqrt{2} & \Delta_B \end{bmatrix} \quad (8)$$

This form of the Hamiltonian shows clearly that when $\Omega_{Ro} \ll \Delta_B$, the coupling to the state $|C_2\rangle$ can be ignored. As such, the collective excitation process leads to a Rabi oscillation in an *effectively closed two level system* consisting of $|A\rangle$ and $|C_1\rangle$. This is the LSIIB in the context of ensemble excitation, and is the key result of this paper.

While it is may be rather obvious at this point, we emphasize nonetheless that we can now represent a quantum bit by this effectively closed two level system. In the process, we have also shown how to perform an arbitrary single qubit rotation, an essential pre-requisite for quantum computing. The details of how such a qubit can be used for quantum computation, quantum communication, and the realization of a quantum internet is described in reference 7.

The essence of the LSIIB for ensemble excitation is summarized in **Error! Reference source not found.**. Briefly, whenever we have a three level optically off-resonant transition for the individual atoms, this can be translated into a corresponding off-resonant three-level transition involving collective states, which in turn is reduced to an effective two-level transition. In order for this to hold, the primary constraint is that, for the collective states, the Rabi frequency on one leg must be much bigger than the same for the other.

To summarize, we have described a new type of blockade that allows one to treat an ensemble excitation as a single, deterministic quantum bit consisting of only two levels. Such as system can be used to realize a two-qubit gate (e.g., a C-NOT gate) between two ensemble-based qubits. It can also be used to transport, deterministically, the quantum state of an ensemble qubit from one location to another separated by macroscopic distances, and it can be used to establish a quantum-link between two ensembles-and-cavity based quantum computers.

## References

[1] R.H. Dicke, "Coherence in spontaneous radiation processes", Phys. Rev. 93, 99(1954).

[2] L.M. Duan, M.D. Lukin, J.I. Cirac, and P. Zoller, "Long-distance quantum communication with atomic ensembles and linear optics", Nature 414, 413 (2001).

[3] L.M. Duan, J.I. Cirac, and P. Zoller, "Three-dimensional theory for interaction between atomic ensembles and free-space light", Phys. Rev. A 66, 023818 (2002).

[4] G. K. Brennen, I. H. Deutsch, And P.S. Jessen, "Entangling Dipole-Dipole Interactions For Quantum Logic With Neutral Atoms," Physical Review A, Volume 61, 062309 (2000)

[5] M.D. Lukin, M. Flieschhauer, R. Cote, L.M. Duan, D. Jacksch, J.I, Cirac, and P. Zoller, "Dipole blockade and quantum information processing in mesoscopic atomic ensembles", Phys. Rev. Lett. 87, 037901 (2001).

[6] T. Pellizzari, S.A. Gardiner, J.I. Cirac, and P. Zoller, " Decoherence, continuous observation, and quantum computing: a cavity QED model", Phys. Rev. Lett. 75 3788 (1995).

[7] M.S. Shahriar, G.S. Pati, and K. Salit "Quantum Communication and Computing With Atomic Ensembles Using Light-Shift Imbalance Induced Blockade [*preprint, available at http://lapt.ece.northwestern.edu/files/QCC-Atomic-Ensemble and at* http://arxiv.org/abs/quant-ph/0604121 ]

[8] S. Lloyd, M.S. Shahriar, J.H. Shapiro, and P.R. Hemmer, "Long Distance, Unconditional Teleportation of Atomic States via Complete Bell State Measurements," Phys. Rev. Lett. **87**, 167903 (2001).

[9] M.S. Shahriar, J. Bowers, S. Lloyd, P.R. Hemmer, and P.S. Bhatia, "Cavity Dark State for Quantum Computing,"*Opt. Commun.* **195**, 5-6 (2001)

[10] S. Lloyd, J. Shapiro, F. Wong, P. Kumar, M.S. Shahriar, and H. Yuen, "Infrastructure for the Quantum Internet," ACM SIGCOMM Computer Communication Review (Oct. 2004).

[11] M.S. Shahriar, P.R. Hemmer, M.G. Prentiss, P. Marte, J. Mervis, D.P. Katz, N.P. Bigelow and T. Cai, "Continuous Polarization-Gradient Precooling Assisted Velocity Selective Coherent Population Trapping," *Phys. Rev. A* (*Rapid Comm*.) **48**, R4034(1993).

[12] M.S. Shahriar, P. Hemmer, D.P. Katz, A. Lee and M. Prentiss, "A Three-element Vector Model for the Resonant Raman Interaction," *Phys. Rev. A*. **55**, 2272 (1997).

[13] P. Hemmer, M. Prentiss, M.S. Shahriar and P. Hemmer, "Optical Force on the Raman Dark State in Two Standing Waves," *Optics Communications*, **89**, 335 (1992).

[14] M.S. Shahriar and P. Hemmer, "Direct Excitation of Microwave-Spin Dressed States Using a Laser Excited Resonance Raman Interaction," *Physical Review Letters*, **65**, 1865(1990).